# Thermochromic Metal Halide Perovskite Windows with Ideal Transition Temperatures

*Bryan A. Rosales[1], Janghyun Kim[1], Vincent M. Wheeler[2], Laura E. Crowe[3], Kevin J. Prince[1,4], Mirzo Mirzokarimov[1], Tom Daligault[1], Adam Duell[1], Colin A. Wolden[1,4], Laura T. Schelhas[1], and Lance M. Wheeler[1]\**

[1]National Renewable Energy Laboratory, 15013 Denver West Parkway, Golden, CO 80401, USA.

[2]University of Wisconsin—Stout, 712 Broadway Street South, Menomonie, Wisconsin 54751, USA

[3]Swift Solar, 981 Bing St, San Carlos, CA 94070, USA

[4]Department Chemical and Biological Engineering, Colorado School of Mines, Golden, CO 80401, USA

E-mail: lance.wheeler@nrel.gov

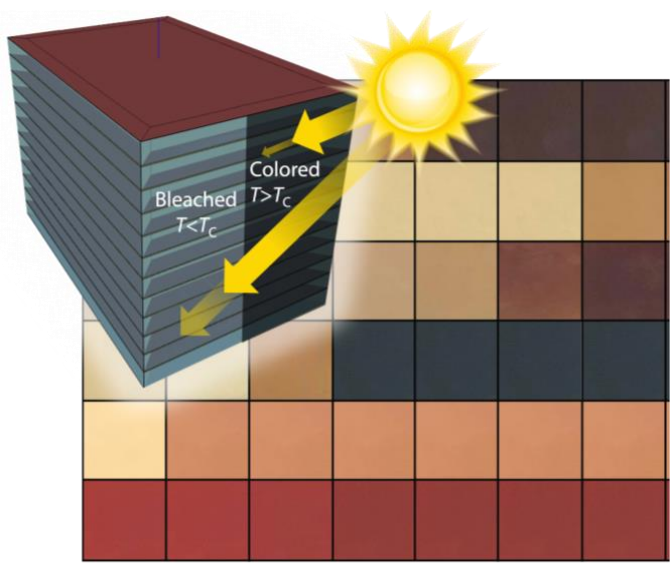

**Abstract**

Urban centers across the globe are responsible for a significant fraction of energy consumption and $CO_2$ emission. As urban centers continue to grow, the popularity of glass as cladding material in urban buildings is an alarming trend. Dynamic windows reduce heating and cooling loads in buildings by passive heating in cold seasons and mitigating solar heat gain in hot seasons. In this work, we develop a mesoscopic building energy model that demonstrates reduced building energy consumption when thermochromic windows are employed. Savings are realized across eight disparate climate zones of the United States. We use the model to determine the ideal critical transition temperature of 20 to 27.5 °C for thermochromic windows based on metal halide perovskite materials. Ideal transition temperatures are realized experimentally in composite metal halide perovskite film composed of perovskite crystals and an adjacent reservoir phase. The transition temperature is controlled by co-intercalating methanol, instead of water, with methylammonium iodide and tailoring the hydrogen-bonding chemistry of the reservoir phase. Thermochromic windows based on metal halide perovskites represent a clear opportunity to mitigate the effects of energy-hungry buildings.

**1. Introduction**

Buildings account for over one-third of the world's final energy consumption and approximately 28% of global $CO_2$ emissions, which increases to >40% when building-related construction is included.[1] Urban areas composed of high-rise buildings continue to gain population, and are predicted to encompass 70% of the world's population by the middle of this century.[2] At the same time, urban skylines increasingly feature glass façades, and the architectural trend across building sectors is toward more glass, despite it greatly underperforming their opaque cladding counterparts for building efficiency.[3]

New window technology must be developed and deployed to reconcile the significant impact buildings have on the environment with the architectural demand for more glazing. Low-emissivity (low-e) coating technology revolutionized window energy-efficiency in the 1980s by selectively absorbing or reflecting infrared wavelengths while maintaining high transmittance of visible light. However, roughly half of the sun's energy is in visible wavelengths. Smart or dynamic window technologies typically target the visible spectrum by dynamically responding to stimulus (light, heat, electric field, etc.) with a change in optical properties. The transition from a visibly absorbing or reflecting (colored) state to a visibly transparent (bleached) state is leveraged to control solar heat gain to either offset heating loads or decrease cooling loads. Electrochromic windows, which transition from bleached to colored states in response to an electric field, offer active control of solar heat gain and glare, but soft costs and installation issues that lead to long payback times have plagued widespread deployment.[4] Thermochromic windows passively respond to changes in temperature to transition optical states. The lack of electrical components bypasses integration and soft-cost issues and enable simpler retrofits and opportunities for widespread deployment.

Thermochromic materials must have an ideal critical transition temperature ($T_C$), fast transition kinetics, a narrow hysteresis width (defined as the difference between the temperatures needed to switch from bleached to colored and colored to bleached), and high solar modulation ability.[5] Thermochromic materials include liquid crystals[6] and leuco dyes,[7] but vanadium dioxide has been established as the quintessential solid-state thermochromic material for building applications. It has been the focus of research for decades due its relatively low-temperature insulator-to-metal Mott transition.[8] Though low compared to most oxides, 68 °C is well above the ideal $T_C$ for window applications. An ideal $T_C$ has been suggested to range between 10 °C and 28 °C based on various reports in the previous decade that typically study simplified buildings and glazing systems (savings reported relative to single-pane windows) and in single climate locations.[5,9,10] Significant research has thus been put into reducing the $T_C$ of $VO_2$ with success in reaching $T_C$<30 °C by using nanostructuring or doping.[11] However, reducing $T_C$ slows the transition kinetics by decreasing the thermodynamic driving force and results in a larger hysteresis width due to the nature of the first-order phase transformation of $VO_2$.

Metal halide perovskite (MHP) materials are a class of semiconductors that have captured the imagination of the materials science community in the last decade due their unmatched optoelectronic properties and scalable solution processibility.[12] Most research has centered on photovoltaics[13] due to their extraordinarily absorption coefficients in the visible and near infrared regions of the solar spectrum. The inherently low formation energy of MHPs[14] enables rapid transformation from the highly absorbing phase to highly transparent ones, which leads to unmatched solar modulation ability. State transformation is induced using intercalation,[15] crystal phase transformation,[16] and nanoparticle precipitation.[17] Each mechanism has now been leveraged to produce thermochromic windows. Perhaps the most interesting feature of MHPs as thermochromic materials is the opportunity to combine chromism with photovoltaic energy generation[15,16,18,19] to bypass the fundamental tradeoff between visible transmittance of a photovoltaic window and power generation.[20] However, ideal transition temperatures for MHP-based thermochromic windows are yet to be demonstrated.

In this work, we combined building energy modeling with detailed materials design of new MHP composite materials for next generation thermochromic windows. We developed a mesoscopic modeling method to investigate building energy savings across eight disparate climate zones of the United States and compare to a baseline of window configurations that meet or exceed current energy efficiency standards. We find thermochromic windows improve energy efficiency across all climate zones and reveal a narrow window for the ideal transition temperature between 20 and 27.5 °C, regardless of climate zone. Our modeling motivates our experimental work to decrease MHP $T_C$ below current demonstration. Incorporation of polymers, chloride, and methanol into the perovskite film introduces a hydrogen-bonding chemistry that enables control over MHP $T_C$. We demonstrate MHP films with $T_C$ tunable down to <22 °C. The films exhibit high solar modulation ability and durability >200 colored-to-bleached cycles due to polymer-induced nanoporous morphology.

## 2. Results and Discussion
### 2.1. Mesoscale modeling determines energy savings and ideal transition temperature ($T_c$)

We developed a mesoscopic building energy analysis model to accurately estimate building energy use across the various climate zones of the United States (Figure 1a). Thermochromic window laminates based on MHPs are simulated using transfer matrix method (TMM) software (*PVwindow*).[21] The TMM code solves Maxwell's Equations for stacks of thin films, such as the perovskite layer (thickness on the order of optical wavelengths), adjacent to thick "incoherent" layers like glass and laminate polymers (e.g. polyvinyl butyral (PVB)). We design the optical properties of the thermochromic laminate to have a dramatic swing between a visible transmittance (VT) of 73% in the bleached state and 5% in the colored state (Figure 1b). It is noteworthy that VT has many different names in the literature but is defined the same way (See Experimental Section), including visible light transmittance (VLT), average visible transmittance (AVT), and luminous transmittance ($T_{Lum}$).

More conventional materials like $VO_2$, which exhibit a metal-to-insulator transition, are capable of modulating the infrared regions of the spectrum. Visible wavelengths make up roughly 51% of the sun's energy, whereas the infrared (IR) and ultraviolet constitute 46% and 3%, respectively. Visible wavelength modulation is critical for energy savings and occupant glare comfort. Here we design thermochromic insulating glass units (IGUs) with low-e layers to statically modulate solar heat gain in the IR with perovskite materials that dynamically modulate the visible portion of the spectrum.

A building-level energy analysis was performed using a physics-based building energy simulation tool (EnergyPlus[24] and OpenStudio[25] software platforms) by incorporating thermochromic IGUs into a highly glazed office building model. IGUs were produced by incorporating optical data of the thermochromic laminates produced from *PVwindow* into software developed by Lawrence Berkeley National Laboratory[22] to create input files with necessary thermal, thermochromic, and optical data for the building simulation tool. Double-pane and triple-pane IGUs are designed to meet ASHRAE 90.1-2019 standards[23] based on center-of-glass U-factor, VT, and solar heat gain coefficient (SHGC) (Figure 1c). Gas fill (air or argon) and optical properties of the panes of glass (VT, emissivity from low-e coatings) are varied to meet regional standards (Table S1, Table S2). The VT and SHGC of the simulated thermochromic laminates are higher (bleached) or lower (colored) than ASHRAE standard properties while maintaining constant thermal properties (e.g. U-factor). The arithmetic average of the bleached and colored VT and SHGC do satisfy the ASHRAE standard, as is accepted practice for thermochromic windows.

A typical meteorological year (TMY3) weather data,[27] which includes realistic sequences of time-dependent weather observations, is used against the building model to simulate various energy consumptions (e.g., heating, cooling, lighting, etc.) in the building in 15-minute intervals for a year in eight climate zones in the United States: Honolulu, HI; Tucson, AZ; San Diego, CA; Denver, CO; New York, NY; Milwaukee, WI; International Falls, MN; and Fairbanks, AK. The total primary energy use was calculated including both electricity and natural gas consumption. By coupling TMM simulation, which accommodate nanoscopic optical effects, to large-scale building energy modeling, we have developed a new bottom-up approach for evaluating window performance in buildings that spans nano-to-macro length scales. The window model relies solely only on the fundamental properties (complex refractive index) and thickness of the materials incorporated. This strategy enables detailed comparison of building performance based on optical switching without the influence of extraneous variables, a difficult task in experimental settings.

We simulated highly glazed office buildings to determine annual energy savings and ideal $T_C$ of the perovskite-based thermochromic windows (Figure 1d). Relative to double-pane low-e windows, thermochromic double-pane windows improve building energy efficiency for each climate zone (Figure 1e). Energy savings are greater in colder regions (Milwaukee, International Falls, and Fairbanks) compared to hotter regions (Honolulu, Tucson, and San Diego). Thermochromic double-pane windows outperform even triple-pane windows in the hottest climate zones (Honolulu), despite their lower U-factors. In colder climates, triple-pane windows provide more energy savings than the thermochromic double-pane windows, but thermochromic triple-pane windows (triple-pane insulating glass unit where the outside pane is exchanged for a thermochromic laminate) provide the most annual energy savings compared to double-pane windows that meet ASHRAE standards.

The source of improved energy efficiency in thermochromic windows, double or triple-pane, is modulation of heating and cooling loads in the building. In cooling dominated climates (Honolulu, Tucson, and San Diego), thermochromic windows reduce cooling load by mitigating solar heat gain (Figure 1e). Heating loads are also reduced in these cases, though less so. An analogous trend is observed for heating-dominated climates (Milwaukee, International Falls, and Fairbanks). Cooling loads notably decrease, but the reduction in heating load during the colder months shows a large benefit due to increased solar heat gain relative to double-pane windows that meet standards for the region, since the windows allow more sunlight in while in the bleached state than low-e double pane windows while still meeting ASHRAE standards. Climates with similar heating and cooling seasons (Denver and New York) benefit from reduced cooling and heating loads with thermochromic windows. Our modeling studies show the significant heating savings potential of thermochromic windows. It is also worth noting thermochromic windows offer the distinct advantage of inexpensive retrofitting of storm windows. Retrofitting the existing building stock for energy efficiency is critical to reducing building energy use by 2050.[29] Our work suggests retrofitting a thermochromic laminate onto a single-pane or even double-pane window will yield significant savings.

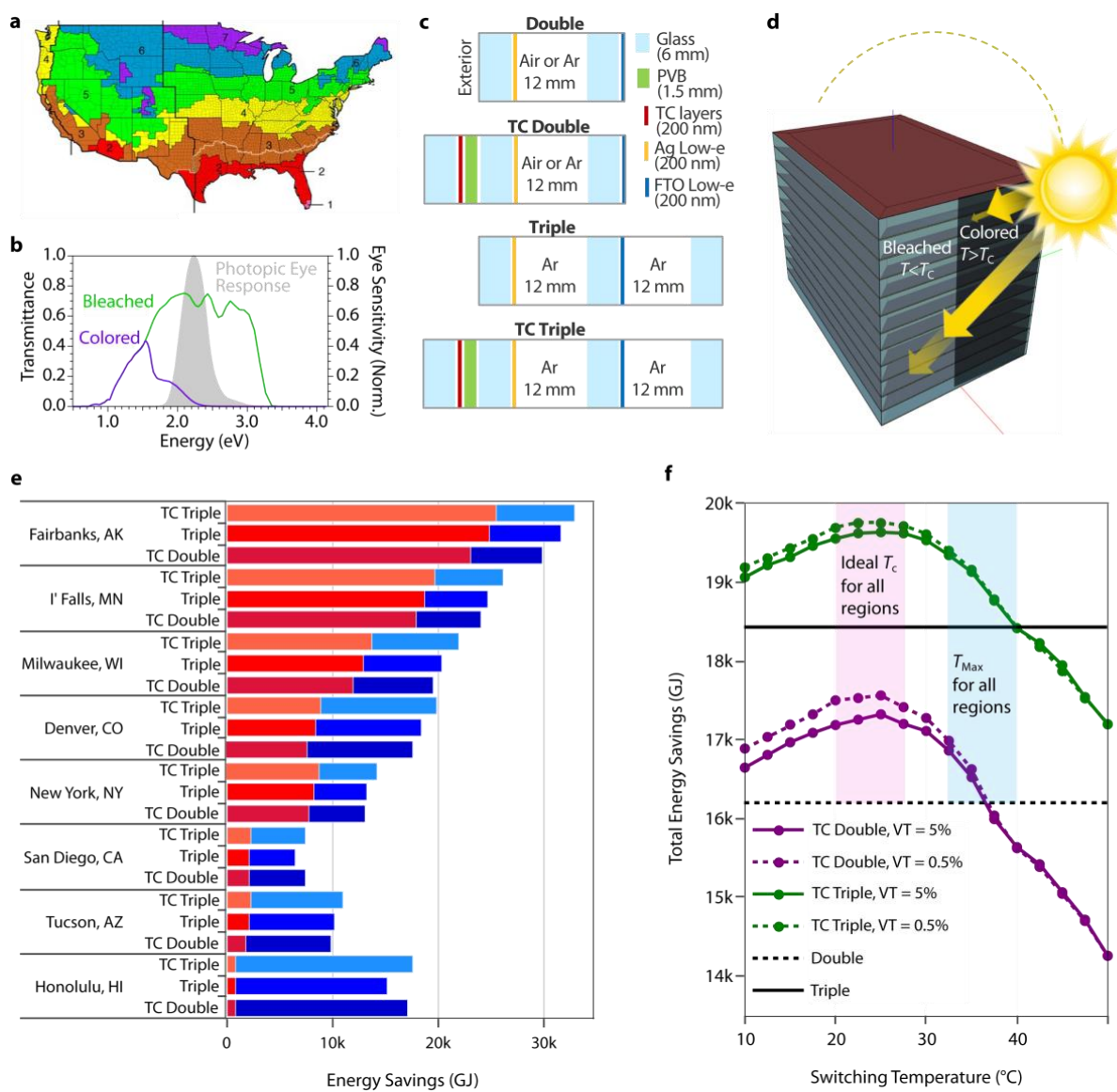

**Figure 1.** (a) Representation of climate zones in the United States, according to the International Energy Conservation Code. Alaska and Hawaii not shown. (b) Transmittance spectra produced by TMM calculation of the thermochromic laminate in colored and bleached states. The gray curve is the photopic eye response to highlight switching in the visible portion of the spectrum. (c) Diagrams of insulating glazing units and (d) highly glazed (95% window-to-wall ratio) 12-story medium office building used for mesoscale building energy analysis. The glass façade transitions from bleached to colored when the glass temperature is greater than the critical transition temperature ($T_C$). PVB = polyvinyl butyral. (e) Annual energy savings due to cooling and heating for eight United States climate zones with varied window construction. We define savings as the absolute difference between buildings with code-compliant glazing with thermochromic layers (TC double, triple-pane, or triple-pane glazing with thermochromic layers (TC Triple), and double-glazed windows with code-compliant low-e coatings that meet current standards for each region of the United States. Cities are ordered from zone 8 (Fairbanks) at the top to zone 1 (Honolulu) at the bottom. (f) Annual energy savings as a function of thermochromic window transition temperature for Denver, CO, USA. Shaded regions represent the range of ideal $T_C$ (pink region) and maximum transition temperatures ($T_{Max}$, blue region) for cities in all climate zones. TC = thermochromic, VT = visible transmittance.

Maximum energy savings is determined by first calculating the thermochromic transition temperature that yields the highest energy savings in each climate zone. Each climate zone exhibits an ideal $T_C$, which maximizes energy savings, as well as a maximum transition ($T_{Max}$) that provides the same annual performance as the static double or triple-pane counterparts (Figure 1g). Transition temperatures exceeding $T_{Max}$ will reduce building efficiency—the building will consume more energy compared to having efficiency-standard-compliant double-pane windows. For example, the ideal $T_C$ is 25 °C, and $T_{Max}$ is 35 °C in Denver. Ideal $T_C$ and $T_{Max}$ do not vary significantly across climate zones, and the range is quite narrow. Ideal $T_C$ is between 20 and 27.5 °C (pink shaded region, Figure 1g), and $T_{Max}$ is between 32.5 and 40 °C (light blue shaded region, Figure 1g). Precise control of the thermochromic transition temperature to reach the ideal $T_C$ is critical to successful deployment of thermochromic windows.

## 2.2. Polymers and Cl enable control of MHP critical transition temperature ($T_C$)

Previous work has demonstrated a tunable $T_C$ of MHP materials to between 35-55 °C depending on the ambient humidity,[30] which is above the ideal switching temperature determined by the modeling described above. Here we tune $T_C$ by formulating MHP-based composites that incorporate chloride and polymers (Figure 2a). We investigate poly(ethylene glycol) (PEG), poly(vinyl alcohol) (PVA), and polyacrylic acid (PAA), which have distinct functional groups that interact with components of the composite (Figure 2b). Switching is enabled by reversible hydration or methanolation to transform the three-dimensionally connected lead halide lattice of the colored state into isolated $[PbI_6]^{4-}$ surrounded by excess methylammonium halide (MAX, where X = I$^-$ or Cl$^-$) and water or methanol of the bleached state (Figure 2c).[31] Thermochromic composite MHP (TMHP) films were fabricated by spin-coating a solution of 4:1 MAI:PbI$_2$ in *N,N*-dimethylformamide (DMF) onto glass followed by annealing at 100 °C for 10 min under inert conditions. Films containing Cl were fabricated by spin-coating a solution 6.5:1 MAI:PbCl$_2$ in DMF onto glass followed by annealing at 100 °C for 1 h under inert conditions. We incorporate polymers into MHP composite films by adding 1 molar equivalent of the polymer with respect to its monomer into the solutions described above (see Experimental section for more details).

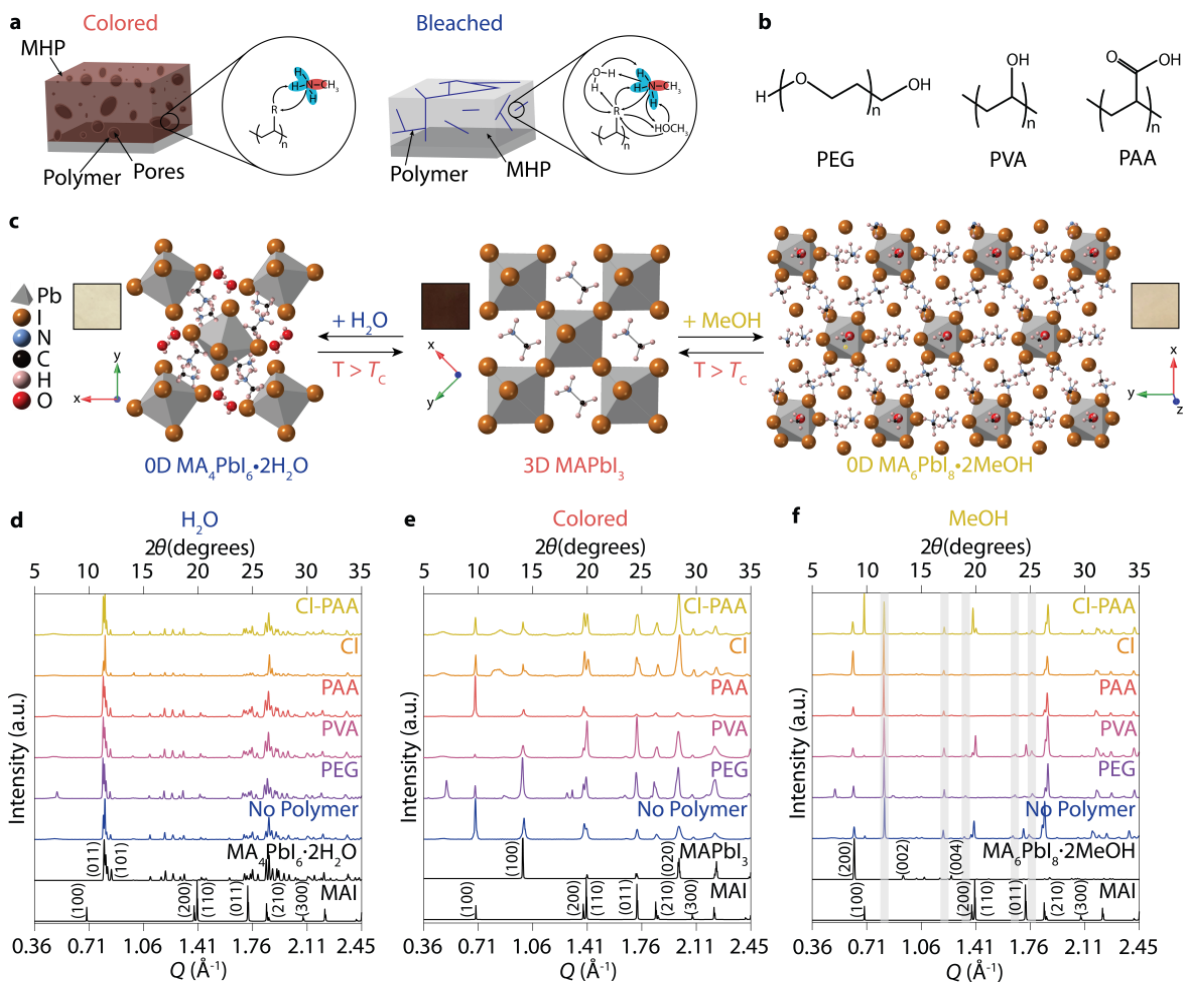

**Figure 2.** (a) Graphical illustration of TMHP films containing polymer in the colored and bleached states with insets illustrating the general molecular interactions present in both states. (b) Structures of the polymers used in this study. (c) Graphical illustration comparing the crystal structures of composite MHP films in the colored phase and upon exposure to $H_2O$ and MeOH. Square insets in (c) are representative photographs comparing the color of each phase. (d-f) Select wide-angle X-ray scattering (WAXS) data extracted from *in-situ* cycling of TMHP films alternatively exposed to $H_2O$/MeOH and 75 °C in Figure S2-S7. Data was obtained from the last WAXS image of the first (b) $H_2O$, (c) heating, and (d) MeOH cycle. Gray bars in (f) indicate the additional phase associated with methanolation. The $2\theta$ axes are relative to Cu K$\alpha$ (1.5406 Å, 8.04 eV) radiation and was calculated from $Q = 4\pi\sin(\theta)/\lambda$ where $\lambda$ is the excitation wavelength.

The crystal structure of the bleached and colored state of the composite films are unaffected by the inclusion of chloride or the polymer (Figures 2d-f). The initial films exhibit a brown color, and wide-angle X-ray scattering (WAXS) shows the (100) and (020) Bragg diffraction peaks of tetragonal MAPbI$_3$ and the (100), (200), (110), (011), (210), and (300) Bragg diffraction peaks of crystalline MAI. The calculated Scherrer crystalline domain size suggests the film is composed of 39 ± 7 nm MAPbI$_3$ nanoparticles surrounded by a "reservoir" of MAI with a domain size of 50 ± 3 nm (Figure 2e).[31, 32] Incorporation of polymers into the composite film does not introduce new diffraction peaks, with the exception of additional crystalline peaks in the PEG TMHP film at 0.50 Å$^{-1}$, 1.29 Å$^{-1}$, and 1.32 Å$^{-1}$. The relative intensity ratios of the MAI peaks change when polymers are introduced, which we attribute to the incorporation of polymer into the "reservoir" of MAI that disrupts the crystal packing of the MAI molecules along different planes. In addition, no new peaks arise due to incorporation of chloride (the extra peak at 0.83 Å$^{-1}$ is due to residual hydrated phases) because MACl is known to evaporate from MAPbI$_3$ films during annealing to leave a small amount of Cl in the film.[30] PAA decreases the relative intensity of the MAI when Cl is incorporated into composite film, which is consistent with PAA incorporation into the reservoir phase.

WAXS shows Bragg diffraction peaks consistent with composite MHP films without polymer in the colored state and upon bleaching with H$_2$O or MeOH, which indicates each thermochromic mechanism is the same (Figure 2d-f). Exposure to H$_2$O or MeOH vapor induces the disappearance of MAPbI$_3$ and MAI Bragg diffraction peaks with the emergence of (011) and (101) peaks characteristic of MA$_4$PbI$_6$•2H$_2$O (Figure 1d), the (200), (002), and (004) peaks characteristic of MA$_6$PbI$_8$•2MeOH (Figure 2f), and additional peaks at 0.82 Å$^{-1}$, 1.21 Å$^{-1}$, 1.34 Å$^{-1}$, 1.65 Å$^{-1}$, and 1.78 Å$^{-1}$ that are associated with methanolation (Figure 2f, gray boxes).[31] Bragg diffraction peaks characteristic of hydration and methanolation disappear when the TMHP film is heated above $T_C$ (Figure 2e).

We determine $T_C$ by heating bleached TMHP films until a color change is observed. There are clear trends upon polymer and Cl incorporation with H$_2$O (Figure 3a) or MeOH (Figure 3b) used as the intercalating molecule. Exposure of TMHP films to H$_2$O vapor at a concentration > 35% relative humidity (RH) bleaches the film due to the structural transformation of 3D MAPbI$_3$ to 0D MA$_4$PbI$_6$•2H$_2$O upon intercalation of H$_2$O. Heating the film above the $T_C$ of hydrated TMHP films (70-75 °C, Figure 3c) reproduces the original brown color (Figure 3a) by dehydrating the film resulting in conversion of 0D MA$_4$PbI$_6$•2H$_2$O back into 3D MAPbI$_3$ (Figure 2e). Similarly, TMHP films incorporating polymer bleach upon exposure to > 35% RH and then darken upon heating above $T_C$ (Figure 3a). We find that $T_C$ of hydrated TMHP films follows the trend PEG (75-80 °C) > No Polymer (70-75 °C) > PVA (65-70 °C) > PAA (60-65 °C). Interestingly, even though incorporation of polymers affects $T_C$, they do not affect the time for bleaching to occur ($t_{bleach}$) upon hydration with all TMHP films exhibiting $t_{bleach}$ < 15 s (Figure 3d).

Exposing the same TMHP films to MeOH vapor also bleaches the film (Figure 3b) due to the structural transformation of 3D MAPbI$_3$ into 0D MA$_6$PbI$_8$•2MeOH and the additional phase (Figure 2f) upon intercalation of MeOH. Heating above the $T_C$ of methanolated TMHP films (45-50 °C, Figure 3c) reproduces the original brown color (Figure 3b) by demethanolating the film resulting in conversion of 0D MA$_6$PbI$_8$•2MeOH and the additional phase back into 3D MAPbI$_3$ (Figure 2f). We find that the $T_C$ of all TMHP films is reduced when MeOH is used as the switching molecule because MeOH exhibits weaker H-bonding compared to H$_2$O. Specifically, the $T_C$ of methanolated TMHP films follows the trend PEG (65-70 °C) > PVA (55-60 °C) > No Polymer (45-50 °C) > PAA (40-45 °C). In contrast to hydration, $t_{bleach}$ varies significantly upon methanolation following the trend PEG (120-130 s) > PVA (90-105 s) > No Polymer (75-90 s) ≅ PAA (75-90 s) (Figure 3d). The higher $t_{bleach}$ of methanolation compared to hydration suggests the driving force for methanolation is weaker than hydration.

Incorporation of a small amount of Cl has been shown to reduce $T_C$.[30] Cl incorporation produces reddish films (Figure 3a and 3b) with a significantly blue-shifted absorbance (Figure S8). The films are bleached upon exposure to both H$_2$O and MeOH vapor with lower $T_C$ values of 40-45 °C for hydrated films and 30-35 °C for methanolated films (Figure 3c). The gradual loss of MACl during film formation also suggests that Cl is influencing the TMHP film formation, and the Cl remaining in the film after annealing suggests that Cl may be facilitating transformations. Combining Cl-doping with the polymer exhibiting the lowest $T_C$, PAA, induces an even further drop in $T_C$ to 30-35 °C for hydrated films and to 20-25 °C for methanolated films (Figure 3c). Cl and Cl-PAA also further reduce $t_{bleach}$ upon methanolation to 60-75 s for both TMHP films.

Incorporation of polymers and Cl allows control of the $T_C$ over a 45 °C or 50 °C window when H$_2$O or MeOH, respectively, is used as the trigger molecule while MeOH allows control over $t_{bleach}$ in a 60 s window. We successfully fabricate a TMHP film exhibiting a $T_C$ within the ideal window of 20-27.5 °C through co-incorporation of Cl with PAA and by using MeOH as the intercalation molecule. Our Cl-PAA films with MeOH exhibit our lowest $T_C$ of 20-25 °C (Figure 3c) and our fastest $t_{bleach}$ upon methanolation of 60-75 s (Figure 3d).

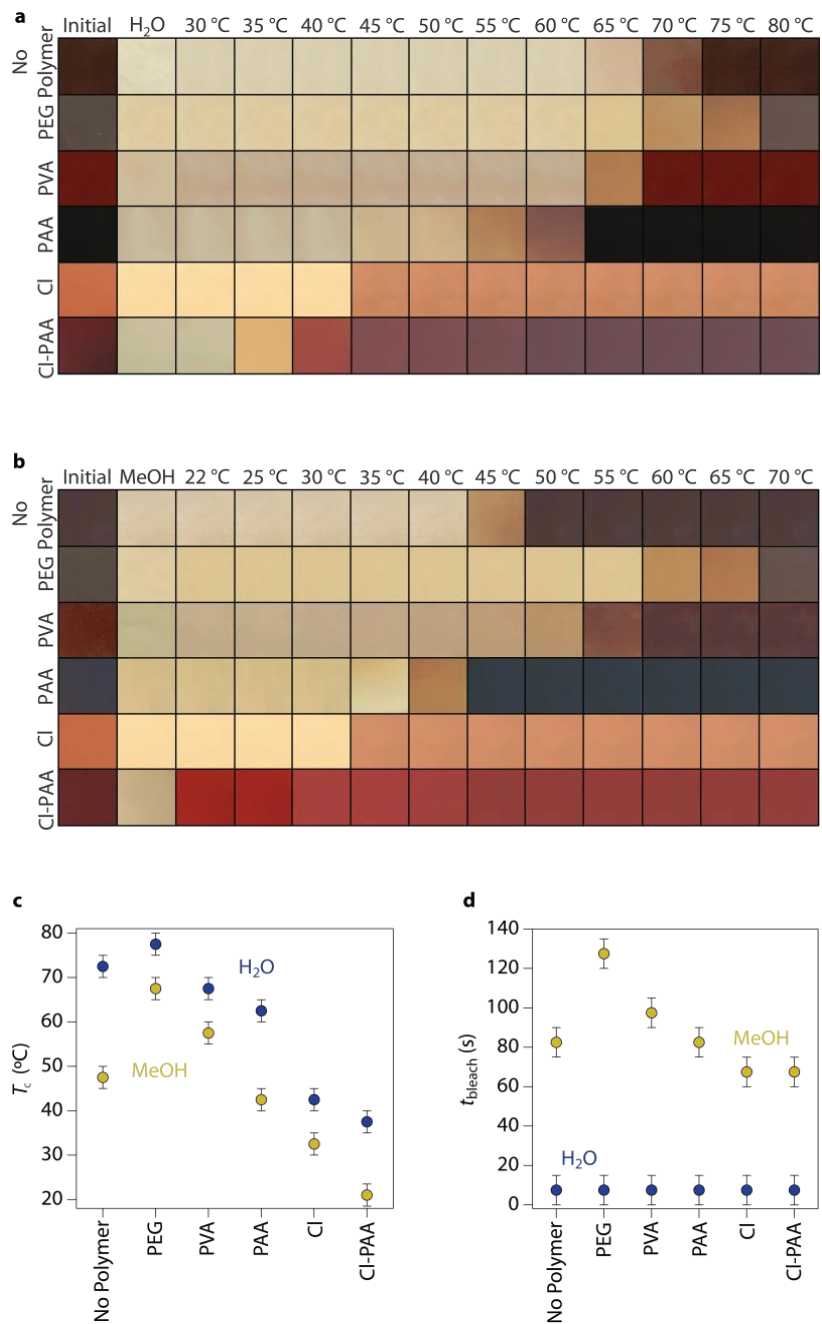

**Figure 3.** Representative photographs showing the color changes of composite MHP films incorporating polymers and/or doped with Cl upon exposure to (a) $H_2O$ and (b) MeOH followed by heating up 70 °C in 5 °C increments to identify the film's $T_C$. (c) Thermochromic transition temperature ($T_C$) for each composite MHP film extracted from (a) and (b). (d) Time for bleaching to occur ($t_{bleach}$) defined as the time for the $MAPbI_3$ (100) peak to disappear as determined from *in-situ* WAXS data with a resolution of 15 s as shown in Figure S2-S7.

**2.3. Tuning switching thermodynamics**

We develop a simple thermodynamic model to describe the observed trends in transition temperature as a function of composite film chemistry. The colored-to-bleached transition occurs due to the intercalant (*i*), composed of MAX•MeOH or MAX•H$_2$O, shuttling from an adjacent reservoir phase (*R*), composed of MAX, MeOH or H$_2$O, and different polymers, into the thermochromic MAPbI$_3$ crystal (*C*) (Figure 4a):

$$iR + C \leftrightharpoons R + iC \qquad (1)$$

The film is colored when the intercalant is thermodynamically favored in the reservoir phase (left side of Eq. 1) and bleached when the intercalant is favored in the crystal phase (Figure 4a). The equilibrium between the bleached state and the colored state is a competition between the binding strength of the intercalant in each host. By changing the Gibbs' energy of the intercalated crystal phase ($G_{iC}$) and the intercalated reservoir phase ($G_{iR}$) we are able to tune which phase is preferred (Figure 4b). A change in Gibbs' energy of the reaction that is greater than zero ($\Delta G > 0$) will yield the bleached state, whereas $\Delta G < 0$ favors the colored phase:

$$\Delta G = G_{iC} - G_{iR} = \begin{cases} >0, \text{ Bleached} \\ <0, \text{ Colored} \\ =0, \text{ Mulitphase} \end{cases} \qquad (2)$$

The thermodynamically stable chromic state is controlled by varying both the strength and number of H-bonds in the system, which will change $G_{iC}$ and $G_{iR}$.[32] For instance, at standard conditions, the bleached state is favored in every chemistry explored in this work except the Cl-PAA sample, which is colored at standard conditions (Figure 3a,b).

$\Delta G_1$ and $\Delta G_2$ are examples of reaction curves where the bleached state is preferred at room temperature (Figure 4b); $G_{iR}$ at a lower energy state than $G_{iC}$. In the simplest system, where no polymer or Cl is included, H-bond interactions between the intercalating species and the metal halide sublattice and MAX reservoir dominate. The energy of the intercalated reservoir state ($G_{iR,1}$) is changed by introducing a polymer or hydrogen bond accepting anion. We show that Cl$^-$ or polymers like PAA increases the strength and number of hydrogen bonds in the reservoir phase, thus reducing $G_{iR,1}$ to $G_{iR,2}$. The energy of the crystal phase ($G_{iC}$) is also controlled with hydrogen bond chemistry. By switching from water to methanol, the hydrogen bond strength in the crystal decreases from $G_{iC,1,2}$ to $G_{iC,3}$ (Figure 4b). $\Delta G_3$ is an example where the colored state is thermodynamically preferred, as in the Cl-PAA sample.

The transition temperature of the reaction is reduced by decreasing $\Delta G$, which is the difference between $G_{iC}$ and $G_{iR}$. Based on the observation of spontaneous reaction to the colored phase at higher temperatures, we conclude the bleached-to-colored reaction is an endothermic process. $\Delta G$ will be negative if the magnitude of the $T\Delta S$ term is greater than $\Delta H$. If the $T\Delta S$ term is less than $\Delta H$, the free energy change will be positive (Figure 4c). In our example, $\Delta G_2 < \Delta G_1$, so $\Delta H_2 < \Delta H_1$ due to a smaller change in the hydrogen bonding environment between the reservoir phase and the crystal phase. If we assume the entropy of the reaction is linear over a small temperature region and does no vary dramatically between $\Delta G_2$ and $\Delta G_1$, then the critical transition temperature is expected to decrease from $T_{C,1}$ to $T_{C,2}$, based on the temperature where $\Delta G = 0$ (Figure 4c).

Attenuated total reflectance Fourier transform infrared spectroscopy (ATR-FTIR) allows us to probe H-bonding at the molecular level. ATR-FTIR spectra of films in the colored phase contain vibrational modes corresponding to MA in MAI and MAPbI$_3$, characterized by vibrational modes of the methylammonium molecules (N-H stretching between 2900-3250 cm$^{-1}$, N-H bending centered at 1557 cm$^{-1}$, N-H rocking centered at 1243 cm$^{-1}$, and C-N stretching centered at 970 cm$^{-1}$) (Figure S9a).[33] Introduction of H$_2$O gives rise to characteristic O-H stretching, and introduction of MeOH exhibits C-H asymmetric stretching, C-H symmetric stretching, and C-O stretching vibrational modes at 3440 cm$^{-1}$, 2970 cm$^{-1}$, 2830 cm$^{-1}$, and 1013 cm$^{-1}$, respectively (Figure S9).[34]

The N-H bond of methylammonium halide molecules is a unique indicator of the H-bonding environment in both colored and bleached states. Here we focus on the N-H bending mode due to its spectral isolation compared to other MA bonds (Figure 4d.) When no polymer is incorporated into the composite MHP film, the vibrational mode is centered at 1557 cm$^{-1}$. Upon exposure to H$_2$O or MeOH, the N-H mode shifts to higher energy, indicating a donation of electron density from the intercalating molecule to methylammonium, which is consistent with typical observations of bonding behavior between amines and hydroxyl groups. The N-H bending peak also broadens, which indicates a more chemically diverse bonding environment due to the presence of the intercalating species.

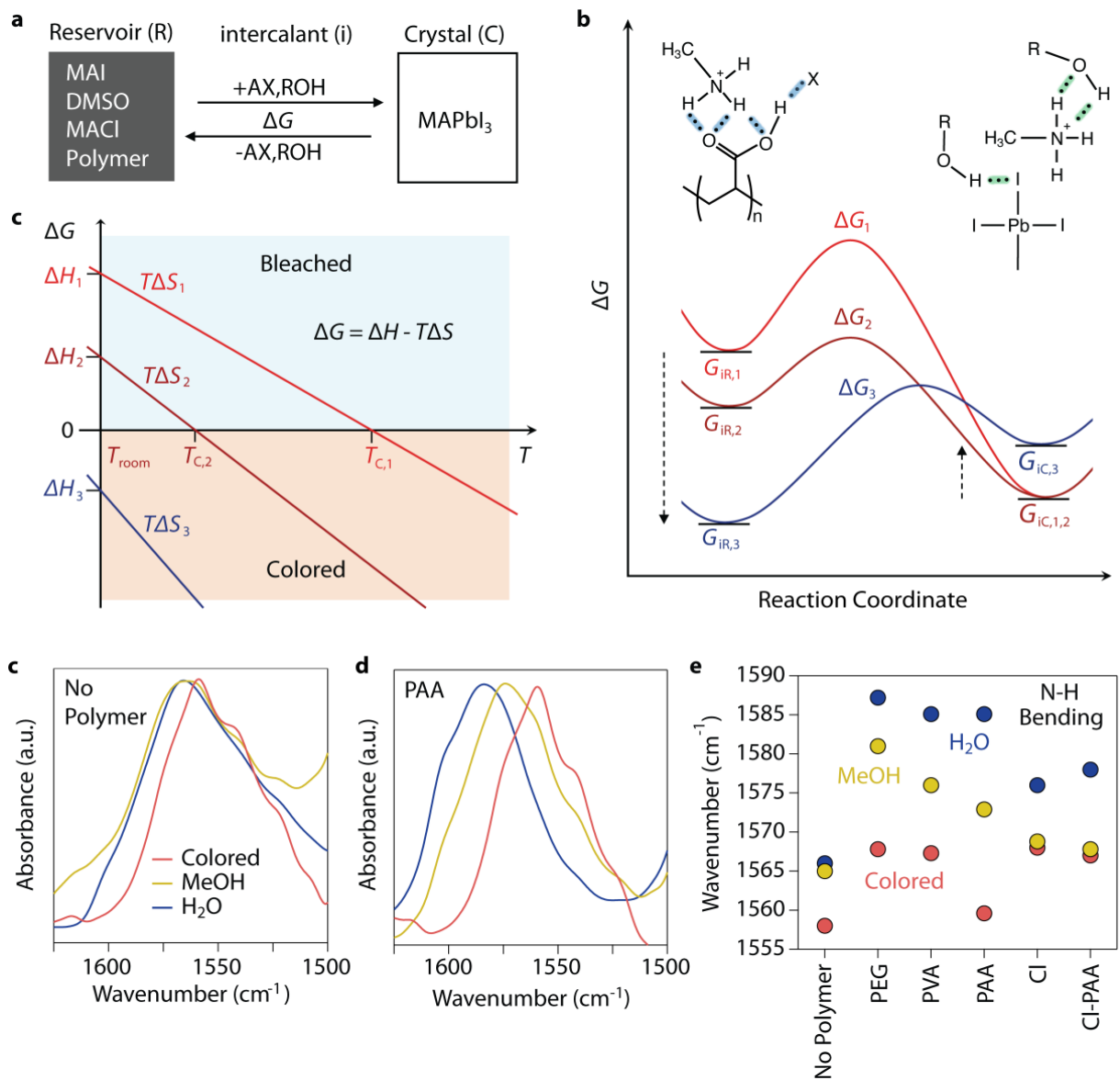

**Figure 4.** (a) Diagram illustrating reaction scheme. (b) Reaction coordinate diagram illustrating control over preferred phase for the intercalant and tunable transition temperature. Representative H-bond interactions in the bleached and color states of the composite TMHP film are shown. R = H or Me. (c,d) Attenuated total reflectance Fourier transform infrared (ATR-FTIR) spectra of the N-H bending mode of TMHP films without polymer (d) and including PAA (e) in the colored state and after exposure to $H_2O$ and MeOH vapor. (e) Peak position of the N-H bending vibrational modes for each TMHP film in the colored state and after exposure to $H_2O$ and MeOH.

The addition of polymers to the composite films is confirmed by ATR-FTIR from the presence of vibrational modes characteristic of the functional groups of the polymers: C-O stretching of PEG at 1098 cm$^{-1}$, O-H stretching of PVA between 3550 cm$^{-1}$ and 3300 cm$^{-1}$, C=O and C-O stretching of PAA at 1710 cm$^{-1}$ and 1170 cm$^{-1}$, respectively. Exposure of the film to $H_2O$ vapor causes characteristic O-H stretching vibrational modes to appear centered at 3490

cm$^{-1}$ (Figure S9).[35] The N-H bending mode signal is a convolution of MA molecules that are in the MHP crystal and those that are in the reservoir. The N-H bending mode is thus a measure of the influence of the polymers incorporated into the reservoir (Figure 4e). For PAA, the N-H bending mode in the colored state blue shifts slightly to 1560 cm$^{-1}$ due to interactions with the carboxylic acid groups of the PAA. MAX•MeOH intercalation broadens and blueshifts the peak by donating and accepting H-bonds with the carboxylic acid groups and the MeOH. H$_2$O further blueshifts the N-H bending peak due to more and stronger H-bonds formed.

The same general blueshifting trend is observed for PEG and PVA (Figure 4d) but to varying degrees. Blue shifting of N-H vibrational modes is comparable between polymers with H$_2$O as the intercalation molecule, whereas blue shifting is increased according to PEG < PVA < PAA with MeOH as the intercalation species. The trend is consistent with the types of H-bonds provided by each polymer. The ether group of PEG will only donate electron density (accept H-bonds), which results in blueshifting of the N-H bond. In contrast, the acidic protons of in the hydroxyl and carboxylic acid groups of PVA and PAA, respectively, will accept and donate H-bonds. The ATR-FTIR signal of the N-H is a convolution of all H-bonds formed within the system, which results in a smaller overall shift due to also donating H-bonds to the polymer.

The trend in $T_C$ between polymers is consistent with the polymer's ability to form H-bonds and effectively increase or decrease $\Delta G$. PEG consistently increases $T_C$ relative to the TMHP film without polymers, and PEG is the weakest H-bonding polymer. PAA, on the other hand, consistently decreases $T_C$ relative to other films. The carboxylic acid groups of PAA form the strongest and highest number of bonds compared to the others, which results in the lowest $T_C$. PVA is in between PEG and PAA by decreasing $T_C$ with H$_2$O and increasing $T_C$ with MeOH relative to the TMHP films without polymer. PVA's hydroxyl group is also capable accepting and donated H-bonds, though the bonds are weaker than those formed with carbonyl groups of carboxylic acid.[36]

The influence of Cl$^-$ is not obvious from the N-H bond signal in ATR-FTIR. The N-H bending mode blueshifts in the presence of Cl$^-$ to a smaller degree than other samples. However, MACl is highly hygroscopic, whereas MAI is not. Cl$^-$ is a hard Lewis base that will more readily accept H-bonds than I$^-$. Cl$^-$ will provide stronger H-bond interactions within the reservoir and may intercalate into the MHP film when bleaching. Stronger bonds to the halide anion in the reservoir phase leads to increased $G_{iR}$ and decreased $T_C$.

## 2.4. Nanoporous TMHP film morphology improves cyclability of TMHPs

The polymers used in this study contain 11k-130k monomers connected in long chains with functional groups capable of H-bonding with the MAX reservoir. These long-chain polymers induce the formation of pores throughout the TMHP film (Figure 5a, colored), and exposing the film to an intercalation molecule leads to a significant decrease in pore density (Figure 5a bleached). The overall thickness of the film is maintained within error during the transformation (colored 250 ± 20 nm; bleached 240 ± 7 nm), which suggests that any volume expansion that may result upon intercalation occurs within the void space of the pores rather than increasing the overall thickness of the film. Heating the film above $T_C$ re-forms the pores while maintaining the film thickness. These observations suggest that the polymers are likely located around the pores in the colored state and at grain boundaries in the bleached state (Figure 2a).

In addition to having a low $T_C$ and rapid switching time, smart windows need to be durable. Cyclability of TMHP films is currently limited by delamination and film reorganization upon repeated intercalation/de-intercalation[15] as well as deprotonation of MAI upon prolonged exposure to $H_2O$ (Figure S10). Polymers have been shown in impart mechanical stability in MHP films and can induce self-healing upon exposure to extreme mechanical stress.[37] Our TMHP films containing polymer are porous with polymer located at the TMHP/pore interface and that volume expansion occurs inward through the pores rather than outward by increasing the film thickness. TMHP films with this morphology should have improved mechanical stability to delamination and cracking. In addition, TMHP films that switch with MeOH exhibit a small driving force for MAI deprotonation due to the higher acidity of MeOH vapor compared to $H_2O$.[31]

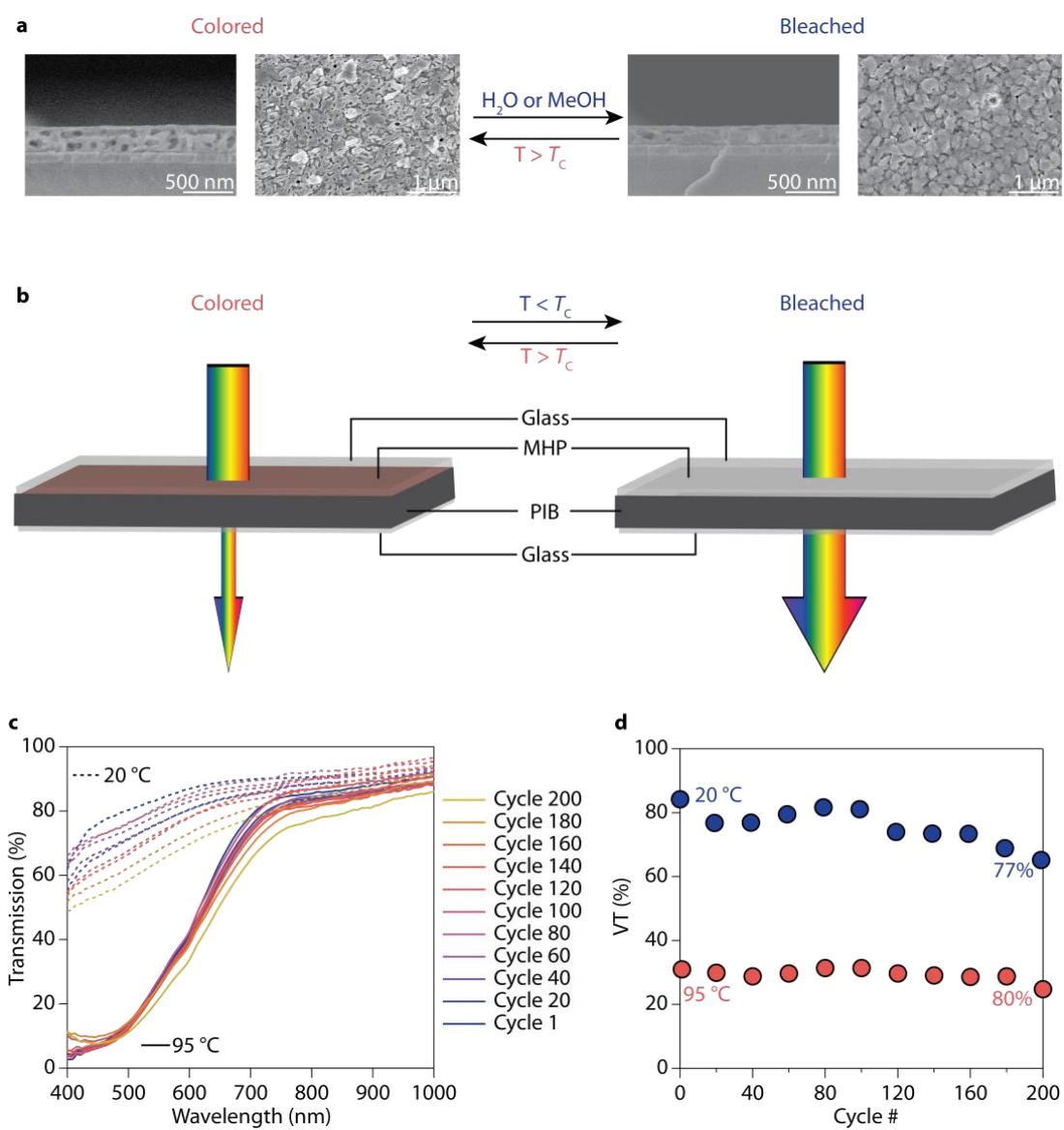

**Figure 5.** (a) Representative scanning electron microscopy (SEM) images of TMHP films containing polymer in the colored and bleached states. (b) Illustration of TMHP smart windows in the colored and bleached states. (c) Transmittance and (d) visible transmittance (VT) data over 200 thermochromic switching cycles of alternating exposure to 20 °C and 95 °C for a TMHP film containing PAA and using MeOH as the intercalation molecule. We note that 95 °C was chosen to increase the thermochromic switching speed.

We fabricated smart windows with TMHP films containing PAA by sealing the film within two pieces of glass containing an atmosphere of $N_2$/MeOH with polyisobutylene (PIB) sealing the edges (Figure 5b). Smart windows with this architecture become colored when heated above $T_C$ and bleached when cooled below $T_C$. The visible light transmittance (VT) of our smart windows cycle between 31% in the colored state and 84% in the bleached state (Figure 5c-d). Remarkably, our smart windows retain 77% of the initial VT in the bleached state and 80% in the colored state over 200 cycles. This result is the most cycles reported in TMHP films to date.

## 3. Conclusion

In this work, we developed a mesoscopic building energy model to evaluate the energy savings and ideal thermochromic switching temperature of MHP-based windows. Our model indicates an increase in building performance across all climate zones. Thermochromic windows supplied a significant savings compared to windows that meet current standards. We showed the most savings in heating dominated climates due to solar heating in winter months. The model also allowed us to determine the ideal thermochromic critical transition temperature, which exists in a tight range of 20 to 27.5 °C. We used the model to motivate our experimental work to tailor metal halide perovskite films to switch at lower temperatures than currently realized. We tuned switching in MHP films down to < 22 °C by employing methanol as the intercalating molecule, including polyacrylic acid in the MAX reservoir, and incorporating chloride anions into the composite. Films demonstrated durable thermochromic switching for 200 bleached-to-colored cycles. The work demonstrates the extraordinary promise of thermochromic metal halide perovskite to reduce building energy consumption and mitigate climate change without sacrificing architectural freedom.

## 4. Experimental Section

*Mesoscopic building energy modeling:* We developed *PVwindow* software in order to simulate nanoscale stacks of materials that compose dynamic or PV glazing (github.com/NREL/PVwindow). The software solves Maxwell's equations for stacks of optically thin and thick materials to yield spectral absorptivity, reflectivity, and transmissivity at a chosen incident solar angle.[21] Simulated spectra are imported into OPTIC 6 software to input the files into the International Glazing Database (IDGB). Files were imported into WINDOW 7 software to simulate insulating glass units, and IDF files were exported for importing into Open Studio[25] for EnergyPlus[24] simulations by incorporating IGUs (installed on all cardinal directions) into the building model. A 12-story medium office building model with 95% window-to-wall ratio was created based on the DOE prototype building model[26] and all the remaining configurations (e.g., wall/roof insulation, equipment efficiency, occupancy schedule etc.) of the building followed the ASHRAE 90.1-2010 standards.[23] A typical meteorological year (TMY3) weather data,[27] which includes realistic sequences of time dependent weather observations, is used against the building model to simulate various energy consumptions (e.g., heating, cooling, lighting, etc.) in the building in every 15-minute intervals and for a year.

*Materials and Chemicals*: $PbI_2$ (99.99%) was purchased from TCI Chemicals; methylammonium iodide (MAI, 99.99%) from Greatcell Solar Materials; *N,N*-Dimethylformamide (DMF, 99.8%, anhydrous), $PbCl_2$ (99.999%), $Al_2O_3$ nanoparticles (<50 nm, 20wt% in IPA), poly(acrylic acid) (PAA, $M_n$ = 130k), poly(ethylene glycol) (PEG, $M_n$ = 30k) from Sigma-Aldrich; poly(vinyl alcohol) (PVA, 98-99% hydrolyzed, $M_w$ = 11k-31k) from Alfa Aesar; methanol (MeOH) from Fisher. All chemicals were used as received.

*Film fabrication and transition temperature determination*: Thermochromic MHP films were fabricated by spinning a 0.2 M-0.6 M solution containing 4:1 MAI: $PbI_2$ (No Polymer TMHP film) or 4:1:1 MAI:$PbI_2$:polymer (PEG, PVA, PAA TMHP films, molar mass of the polymer's monomer was used) in DMF on glass at 4000 rpm for 30s followed by annealing at 100 °C for 10 min under $N_2$. Films containing Cl were fabricated by the same methods except a solution containing 6.5:1 MAI:$PbCl_2$ (Cl TMHP film) or 6.5:1:1 MAI:$PbCl_2$:PAA (Cl-PAA TMHP film) in DMF was used instead and annealing was performed at 100 °C for 1 h under $N_2$. $H_2O$ vapor was introduced *via* a glovebox equipped with a humidity controller connected to a humidity sensor, commercial humidifier, and house compressed air as a dehumidifier. MeOH vapor was introduced by placing a TMHP film into a container with liquid MeOH at the bottom until the TMHP film turns transparent and colorless. Overexposure to MeOH causes the TMHP film to turn opaque white with reduced thermochromic response. The thermochromic transition temperature ($T_C$) was measured by heating a bleached TMHP film on a hotplate kept at < 15% RH at 5 °C intervals measured by an infrared thermometer gun until the TMHP film transitions from colorless to a dark.

*Structural and morphological characterization. In situ* wide-angle X-ray diffraction (WAXS) data was collected at the Stanford Synchrotron Radiation Light Source (SSRL) at beamline 11-3. The TMHP film were measured at an incident angle of 3 degrees and an incident X-ray wavelength of $\lambda_{synchrotron}$ = 0.9744 Å. A Rayonix MX225 2D detector was used to collect data, and a $LaB_6$ standard used to calibrate the data. MeOH vapor was introduced to the sample chamber by flowing He through a bubbler containing MeOH *via* a Schlenk line. $H_2O$ vapor was introduced to the sample chamber using a commercial room humidifier powered by a humidity controller, which was connected to a humidity sensor within the sample chamber. MeOH and $H_2O$ vapor concentrations in the sample chamber were reduced by flowing helium gas through the chamber. Data acquisition was continuous except for brief interruptions for changing the sample chamber connection between MeOH, $H_2O$, and He. The integration time per measurement was 15 s. The data was integrated using GSASII.[38] $Q$ and $2\theta$ values were calculated by converting $2\theta_{synchrotron}$ to $Q$ ($Q = 4\pi\sin(2\theta/2)/\lambda_{synchrotron}$) and to $2\theta_{Cu\ K\alpha}$ ($2\theta_{Cu\ K\alpha} = 2*\arcsin(Q*\lambda_{Cu\ K\alpha}/4\pi)$ 2) relative to Cu Kα (1.5406 Å, 8.04 eV). Scherrer analysis was performed using a κ value of 0.9 and FWHM values with error bars obtained by fitting peaks to a Voigt function. Scanning electron microscopy (SEM) images were collected with a Hitachi S-4800 Field Emission SEM and TMHP film were prepared on ITO substrates.

*Optical characterization.* Absorbance of films and transmittance of smart windows was collected with an Ocean Optics Maya 2000 Pro UV-Vis Spectrophotometer. Films were prepared as above except No Polymer and Cl TMHP films were spun on a ~400 nm thick mesoporous $Al_2O_3$ scaffold prepared by spinning a 5wt% solution of $Al_2O_3$ NPs in IPA onto UV-ozoned glass at 3000 rpm for 30s followed by annealing at 500 °C for 30 min. All spectra were collected at 22 °C and substrate absorbance was subtracted from all spectra. Visible transmittance is determined by the number of photons transmitted through window, T, weighted by the sensitivity of the human eye to see those photons, which we define as the CIE photopic luminosity function, $\varphi = \varphi(E)$[24]:

$$VT = \frac{\int \Gamma T \varphi \, dE}{\int \Gamma \varphi \, dE}$$

The value is divided by the total incident intensity, $\Gamma$, weighted by $\varphi$ and integrated over the total spectrum giving values between zero and one.

*ATR-FTIR measurements.* A Bruker Alpha FTIR spectrometer outfitted with a diamond attenuated total reflection Fourier transform infrared (ATR-FTIR) spectroscopy attachment with heating and cooling capabilities was used in the study. TMHP films were made by drop-casting 1 μL of the above solutions with a concentration of 0.1 M (No Polymer and Cl films) or 0.05 M (PEG, PVA, PAA, Cl-PAA films) onto the ATR crystal. A custom glass chamber with an O-ring was placed over the ATR crystal stage to enclose the solution and connect the stage to a Schlenk line. The solution was annealed at 100 °C for 30 min under $N_2$ flow to ensure all DMF was evaporated prior to data acquisition. The film was bleached by flowing $N_2$ through a bubbler containing MeOH or $H_2O$ at a rate of 1-3 bubbles per s at 25 °C. Flowing $N_2$ was used to purge the glass chamber encasing the film before data acquisition to ensure that only MeOH or H$_2$O within the film and not in the atmosphere was detected. MeOH or H$_2$O were removed from the film by heating the ATR stage to 100 °C and the stage was cooled back to 25 °C before data acquisition. The temperatures reported are those measured and delivered to the ATR stage with OPUS 7.2 software. All spectra were collected between 350 and 4000 cm−1 with a resolution of 2 cm$^{-1}$ at 25 °C under N$_2$ flow.

*Smart window fabrication.* Thermochromic MHP films for smart windows were fabricated as described above with a 0.2 M solution containing 4:1:1 MAI:PbI$_2$:PAA in DMF. Thermochromic MHP on the outer 3 mm edge of the glass substrate was scraped off with a razor blade. A layer of PIB was placed on the clean edge, another glass substrate was placed on top, and the glass layers were pressed together to form a seal. The atmosphere inside the smart window was replaced with N$_2$/MeOH or N$_2$/H$_2$O by poking a hole in the PIB, flowing N$_2$ through a bubbler containing MeOH or H$_2$O at a rate of 1-2 bubbles per s until the film transitions from colored to transparent colorless, and then immediately plugging the hole with more PIB.

**Supporting Information**

Supporting Information includes: thermochromic laminate diagram, tables of IGU properties, WAXS data, Absorbance spectra of TMHP films, and ATR-FTIR spectra of TMHP films.

**Conflicts of Interest**

The authors declare no conflict of interest.

**Acknowledgements**

This study was authored by the National Renewable Energy Laboratory, operated by Alliance for Sustainable Energy, LLC, for the U.S. Department of Energy (DOE) under contract No. DE-AC36-08GO28308. Funding was provided by the Building Technologies Office within the U.S. Department of Energy Office of Energy Efficiency and Renewable Energy. Use of the Stanford Synchrotron Radiation Lightsource, SLAC National Accelerator Laboratory, was supported by the U.S. Department of Energy, Office of Basic Energy Sciences under Contract No. DE-AC02-76SF00515. The views expressed in the article do not necessarily represent the views of the DOE or the U.S. Government. The U.S. Government retains and the publisher, by accepting the article for publication, acknowledges that the U.S. Government retains a nonexclusive, paid-up, irrevocable, worldwide license to publish or reproduce the published form of this study, or allow others to do so, for U.S. Government purposes.**References**

(1) Tracking Buildings 2020. https://www.iea.org/reports/tracking-buildings-2020 (accessed Sept. 26th, 2021).
(2) Seto, K. C. e. a. *Human Settlements, Infrastructure and Spatial Planning*. Climate Change 2014: Mitigation of Climate Change, Contribution of Working Group III to the Fifth Assessment Report of the Intergovernmental Panel on Climate Change, 2014; pp 923− 1000.
(3) Wilson, A., Rethinking the All-Glass Building: Is It Time to Stop Designing Buildings with Highly Glazed Façades? *Building Green* **2010,** *19* (7), 19.
(4) Needell, D. R.; Phelan, M. E.; Hartlove, J. T.; Atwater, H. A., Solar Power Windows: Connecting Scientific Advances to Market Signals. *Energy* **2021,** *219*, 119567.
(5) Aburas, M.; Soebarto, V.; Williamson, T.; Liang, R.; Ebendorff-Heidepriem, H.; Wu, Y., Thermochromic Smart Window Technologies for Building Application: A Review. *Appl. Energy* **2019,** *255*, 113522.
(6) Seredyuk, M.; Gaspar, A. B.; Ksenofontov, V.; Reiman, S.; Galyametdinov, Y.; Haase, W.; Rentschler, E.; Gütlich, P., Room Temperature Operational Thermochromic Liquid Crystals. *Chem. Mater.* **2006,** *18* (10), 2513-2519.
(7) Kulčar, R.; Friškovec, M.; Hauptman, N.; Vesel, A.; Gunde, M. K., Colorimetric Properties of Reversible Thermochromic
Printing Inks. *Dyes Pigm.* **2010,** *86* (3), 271-277.
(8) Chang, T.-C.; Cao, X.; Bao, S.-H.; Ji, S.-D.; Luo, H.-J.; Jin, P., Review on Thermochromic Vanadium Dioxide Based Smart Coatings: From Lab to Commercial Application. *Adv. Manuf.* **2018,** *6* (1), 1-19.
(9) Long, L.; Ye, H., How to be Smart and Energy Efficient: A General Discussion on Thermochromic Windows. *Sci. Rep.* **2014,** *4*, 6427.
(10) Hong, X.; Shi, F.; Wang, S.; Yang, X.; Yang, Y., Multi-Objective Optimization of Thermochromic Glazing Based on Daylight and Energy Performance Evaluation. *Build. Simul.* **2021,** *14* (6), 1685-1695.
(11) Warwick, M. E. A.; Binions, R., Advances in Thermochromic Vanadium Dioxide Films. *J. Mater. Chem. A* **2014,** *2* (10), 3275-3292.
(12) Dou, B.; Whitaker, J. B.; Bruening, K.; Moore, D. T.; Wheeler, L. M.; Ryter, J.; Breslin, N. J.; Berry, J. J.; Garner, S. M.; Barnes, F. S.; Shaheen, S. E.; Tassone, C. J.; Zhu, K.; van Hest, M. F. A. M., Roll-to-Roll Printing of Perovskite Solar Cells. *ACS Energy Letters* **2018,** *3* (10), 2558-2565.
(13) Kim, J. Y.; Lee, J. W.; Jung, H. S.; Shin, H.; Park, N. G., High-Efficiency Perovskite Solar Cells. *Chem. Rev.* **2020,** *120* (15), 7867-7918.
(14) Rosales, B. A.; Schutt, K.; Berry, J. J.; Wheeler, L. M., The Yin and Yang of Formation Energy, Polymorphism, and Ion Transport in Halide Perovskites *Manuscript Under Review*.
(15) Wheeler, L. M.; Moore, D. T.; Ihly, R.; Stanton, N. J.; Miller, E. M.; Tenent, R. C.; Blackburn, J. L.; Neale, N. R., Switchable Photovoltaic Windows Enabled by Reversible23